\documentclass[preprints,article,accept,moreauthors,pdftex,10pt,a4paper]{Definitions/mdpi} 

\firstpage{1} 
\pubvolume{xx}
\issuenum{1}
\articlenumber{5}
\pubyear{2018}
\copyrightyear{2018}
\history{Received: date; Accepted: date; Published: date}

\usepackage{physics}
\usepackage{amssymb}

\newcommand{\ham}{\ensuremath{H}}
\newcommand{\nspin}{\ensuremath{N}}
\newcommand*{\api}[1]{\ensuremath{^{\text{#1}}}}
\newcommand*{\ped}[1]{\ensuremath{_{\text{#1}}}}
\newcommand{\tf}{\ensuremath{t\ped{f}}}
\newcommand{\omegac}{\ensuremath{\omega\ped{c}}}
\newcommand{\rng}[2]{\comm{#1}{#2}}
\DeclareMathOperator{\iu}{i}
\DeclareMathOperator{\eu}{e}
\DeclareMathOperator{\diss}{\mathcal{D}}

\Title{May a dissipative environment be beneficial for quantum annealing?}

\Author{Gianluca Passarelli $^{1, 2,}$*, Giulio De Filippis $^{1, 2}$, Vittorio Cataudella $^{1, 2}$ and Procolo Lucignano $^{1, 2}$}

\AuthorNames{Gianluca Passarelli, Giulio De Filippis, Vittorio Cataudella and Procolo Lucignano}

\address{%
$^{1}$ \quad Dipartimento di Fisica ``E.~Pancini'', Università di Napoli ``Federico II'', Monte S.~Angelo, Via Cinthia, I-80126 Napoli, Italia\\
$^{2}$ \quad CNR-SPIN, Monte S.~Angelo,  via Cinthia,  I-80126 Napoli, Italia}

\corres{Correspondence: gpassarelli@fisica.unina.it}

\abstract{We discuss the quantum annealing of the fully-connected ferromagnetic $ p $-spin model in a dissipative environment at low temperature. This model, in the large $ p $ limit, encodes in its ground state the solution to the Grover's problem of searching in unsorted databases. In the framework of the quantum circuit model, a quantum algorithm is known for this task, providing a quadratic speed-up with respect to its best classical counterpart. This improvement is not recovered in adiabatic quantum computation for an isolated quantum processor. We analyze the same problem in the presence of a low-temperature reservoir, using a Markovian quantum master equation in Lindblad form, and we show that a thermal enhancement is achieved in the presence of a zero temperature environment moderately coupled to the quantum annealer.}


\begin{document}


\section{Introduction}

NP-hard and NP-complete optimization problems can be mapped on Ising spin Hamiltonians, whose ground states encode the solution of the given problem~\cite{lucas}. In adiabatic quantum computers~\cite{harris}, the target ground state is reached via a so-called quantum annealing procedure~\cite{nishimori}. This technique exploits quantum tunneling as an attempt to explore the system phase space more efficiently than classical analogues, such as simulated thermal annealing. In the last decade, many efforts have focused on studying quantitatively how the efficiency of quantum annealing is affected by dissipation, as technological advancements in quantum computation have to take into account the unavoidable interaction of quantum processors with their surroundings. 

We contribute to this topic by analyzing the dissipative quantum annealing of the ferromagnetic $ p $-spin model, a fully-connected Ising system which is known to encode the Grover's search task~\cite{grover} in the limit $ p\to\infty $ (odd $ p $). The environment surrounding the qubit system is modeled as a collection of harmonic oscillators. The approximate dynamics of the spin system is obtained by tracing away the degrees of freedom of the reservoir and using a Markovian quantum master equation in Lindblad form, preserving the complete positivity of the reduced density matrix. Details on the model Hamiltonian, annealing procedure and Lindblad equation are provided in the next section.
 
 \section{Materials and Methods}
 
 Our aim is to study the ferromagnetic $ p $-spin model, whose Hamiltonian is given by ($ \hslash = 1 $ here and in the following)
 \begin{equation}
	 \ham_1 = -E\nspin \qty(\frac{1}{\nspin} \sum_{i = 1}^{\nspin} \sigma_i^z)^p,
 \end{equation}
 where $ \sigma_i^z $ is the Pauli operator of the $ i $\api{th} qubit along the $ z $ direction and $ E $ is an energy scale of the spin-spin interaction. Its ground state is ferromagnetic and non-degenerate if $ p $ is odd. Quantum tunneling is introduced via a transverse field Hamiltonian
 \begin{equation}
	 \ham_0 = -\Gamma\sum_{i = 1}^{\nspin} \sigma_x^i,
 \end{equation}
 where $ \Gamma $ is the strength of the transverse field energy and $ \sigma_x^i $ are Pauli operators along the $ x $ direction. Given an annealing time $ \tf $ and the dimensionless evolution time $ s = t / \tf \in \rng{0}{1} $, we specify an annealing protocol assigning two functions of time $ A(s) $ and $ B(s) $, monotonically decreasing from $ 1 $ to $ 0 $ and increasing from $ 0 $ to $ 1 $, respectively, and we build the time-dependent Hamiltonian $ \ham(s) = A(s) \ham_0 + B(s) \ham_1 $. For simplicity, we will restrict here to the case $ A(s) = 1 - s $ and $ B(s) = s $. The spin symmetry of this Hamiltonian further simplifies numerical simulations: $ \comm{\ham(t)}{S^2} = 0 $.
 
 At $ s = 0 $, the qubit system is initialized in the ground state of $ \ham_0 $. Then, quantum fluctuations are slowly annealed to zero and the system is evolved in time, up to $ s = 1 $. As prescribed by the adiabatic theorem, the system will remain in its instantaneous ground state if the evolution time is slow compared to the time scale set by the inverse minimal gap between the ground state and the first excited state.  This spin system undergoes a quantum phase transition (QPT) at $ A(s^*) \Gamma = B(s^*) E  $, separating its para- and ferromagnetic phases. For $ p > 2 $, this QPT is of first-order and the minimal spectral gap scales as an exponentially vanishing function of the system size in the thermodynamical limit, making the Grover's search limit non-trivial to retrieve using quantum annealing. 
 
 In the presence of a thermal bath, the $ \text{reduced system} + \text{bath} $ Hamiltonian reads $ \ham\ped{SB}(s) = \ham(s) + \ham_B + V $, where $ \ham_B $ is the Hamiltonian of a collection of independent bosons and $ V $ is an interaction potential which couples the qubit system to its environment:
 \begin{equation}
	 \ham_B = \sum_k \omega_k a_k^\dagger a_k; \qquad V = g \sum_{i = 1}^{\nspin} \sigma_i^z \sum_k \qty(a_k + a_k^\dagger). 
 \end{equation}
 In the continuum limit, the spectral function of the environment considered here takes the form
 \begin{equation}
	 J(\omega) = g^2 \sum_k \delta(\omega - \omega_k) = \eta \omega \eu^{-\omega / \omegac},
 \end{equation}
 where $ \eta $ is a dimensionless effective coupling and \omegac{} is a high-frequency cut-off.  
 
In the weak-coupling regime, and assuming that the bath is initialized in a thermal equilibrium state at temperature $ 1 / \beta $ ($ k\ped{B} = 1 $ here and in the following), the Born, Markov and rotating wave approximations allow to derive a dynamical equation for the reduced density matrix which is in the so-called Lindblad form~\cite{zanardi}:
 \begin{equation}
	 \dv{\rho(t)}{t} = \iu \comm{\rho(t)}{H(t) + H\ped{LS}(t)} + \diss[\rho(t)];
 \end{equation}
 this is equivalent to disregarding qubits-bath correlations and memory effects, and to enforcing energy conservation. To quantify the performances of the quantum annealing, we refer to the residual energy per spin at $ t = \tf $, which is an indicator of the error of the annealing procedure:
 \begin{equation}
	 \epsilon = E + \Tr\bigl[\rho(\tf) \ham(\tf)\bigr] / \nspin.
 \end{equation} 
 
\section{Results}

We show our results for a chain of $ \nspin = 16 $ qubits, with $ p = 3 $. This choice is motivated by the fact that $ p = 3 $ is the hardest case for the $ p $-spin model. For $ p > 3 $, the introduction of a non-stoquastic potential in the Hamiltonian is expected to turn the first-order QPT into a second-order QPT, where the gap closes polynomially as a function of the system size. This procedure does not hold for $ p = 3 $~\cite{nishimori:non-stoq}. We measure energies in units $  E $ and times in units $ E^{-1} $, and we set $ \Gamma = E $. With this choice, the minimal gap is equal to $ \Delta \approx 0.47 E  $.  We choose $ \omegac = 50 E $ and $ 1 / \beta = E / 10 $, and we evaluate the residual energy of quantum annealing for several values of the coupling strength $ \eta $, in the range $ \rng{0}{10^{-2}} $, for annealing times ranging from $ \tf = 0 $ to $ \tf = 10^{4} / E $. 

\begin{figure}[H]
	\centering
	\includegraphics[width = 0.6\textwidth]{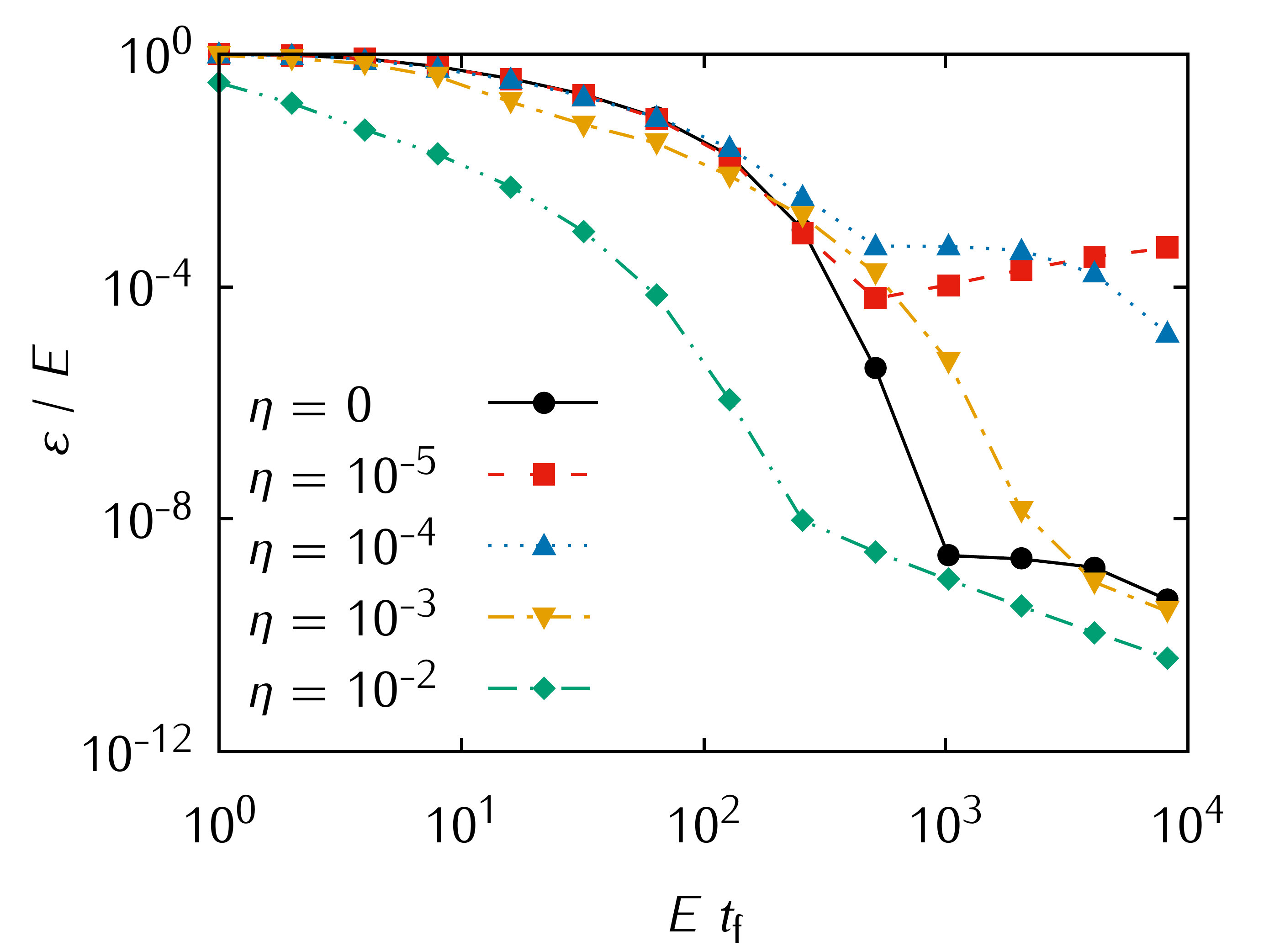}
	\caption{Residual energy as a function of the final annealing time, for $ \nspin = 16 $, $ p = 3 $, $ \Gamma = E $, $ \omegac = 50 E $ and $ \beta = 10 / E $. For the the two smallest coupling strengths analyzed, the bath is always detrimental for quantum annealing. Instead, for stronger couplings, the bath can be beneficial for the annealing.}
	\label{fig:qa-n-16-p-5-beta-10}
\end{figure}

As shown in Figure~\ref{fig:qa-n-16-p-5-beta-10}, for small values of $ \eta $ the bath has a detrimental effect on the reduced system dynamics and the residual energy of the dissipative quantum annealing is higher than the closed one. This is due to the fact that thermal excitations across the gap contribute to reduce the ground state population, even if the thermal energy is lower than $ \Delta $, as multiple-boson processes contribute to this effect. In this parameter region, the dynamics of qubits and bath are uncorrelated and the environment drives the reduced system towards its thermal equilibrium state at temperature $ 1 / \beta $. 

If we increase the coupling strength further, we reach a regime where the environment enhances the performances of the annealing procedure, leading to a faster convergence of the adiabatic algorithm even at short and intermediate times. We interpret this result as the insurgence of an entangled system-bath state which dynamically leads to an effective broadening of the spectral gap of the reduced system, thus providing a thermal enhancement over the isolated case. We claim that this is a quantum effect as it survives also at zero temperature, as confirmed by the recent literature~\cite{passarelli}. The thermal improvement might be useful to increase the fidelity of the annealing procedure even at those times where the adiabatic theorem does not hold. Notice, however, that for this choice of parameters the Lindblad approximation might fail to give the correct results, as effects beyond the Born-Markov approximation begin to play a role at not-so-weak coupling strengths and low temperatures. 

\section{Discussion}

We report the observation of a thermal enhancement for the dissipative quantum annealing of the ferromagnetic $ p $-spin model, whose dynamics has been derived with a commonly used Markovian quantum master equation and discussed for $ \nspin = 16 $ and $ p = 3 $. This effect, observed at intermediate coupling strength and low temperature, is most likely due to the formation of a system-bath entangled state, which helps the reduced system reach its ground state and may be used to retrieve the solution to the Grover's search problem faster than isolated quantum annealing, and to improve convergence at short and intermediate annealing times, i.\,e., when the adiabatic theorem does not hold. As the Lindblad approximation is not completely satisfactory in this parameter region, we are currently extending this work employing an alternative technique accounting for effects beyond the Born-Markov approximation. We successfully applied this method for the spin-boson model~\cite{loris} and we are currently extending it to many-body systems.

%
%
%
%
%


\vspace{6pt} 



\authorcontributions{conceptualization, P.\,L. and G.\,P.; software, P.\,L. and G.\,P.; formal analysis, G.\,P.; investigation, P.\,L., G.\,P., V.\,C. and G.\,D.\,F.; writing—original draft preparation, G.\,P.; writing—review and editing, G.\,P. and P.\,L.; visualization, G.\,P.; supervision, P.\,L., V.\,C. and G.\,D.\,F.}
	

\funding{This research received no external funding.}


\conflictsofinterest{The authors declare no conflict of interest.} 

\reftitle{References}





\end{document}